\documentclass{article}[12pt]
\usepackage{amsmath}
\usepackage{amssymb}
\usepackage{amsthm}
\usepackage{amsfonts}
\usepackage[dvips]{graphicx}

\def\abstract#1{\vskip 7mm
        \begin{center}{\large Abstract}\par \smallskip
                \begin{minipage}[c]{12cm}
                        \small #1
                \end{minipage}
        \end{center}
}

\def\title#1{\begin{center}{\Large\bf #1}\end{center}}
\def\author#1{\vskip 5mm \begin{center}{#1}\end{center}}
\def\address#1{\begin{center}{\it #1}\end{center}}

\newcommand{\ssmatrix}[4]%
{\begin{pmatrix} #1 & #2 \\ #3 & #4 \end{pmatrix}}
\makeatletter
\@ifundefined{lesssim}{}{}
\@ifundefined{gtrsim}{}{}
\def\vereq#1#2{\lower3pt\vbox{\baselineskip1.5pt \lineskip1.5pt
\ialign{$\m@th#1\hfill##\hfil$\crcr#2\crcr\sim\crcr}}}
\makeatother

\newtheorem{theorem}{Theorem}

\newtheorem{proposition}{Proposition}
\newtheorem{remark}{Remark}
\newtheorem{lemma}{Lemma}
\theoremstyle{definition}
\newtheorem*{definition}{Definition}
\theoremstyle{remark}
\newtheorem*{remark*}{Remark}

\newcommand{\vtr}[2]{\paren{\begin{array}{c} #1 \\ #2 \end{array}}}

\newcommand{\EPS}{{\cal E}}
\newcommand{\eh}{event horizon}
\newcommand{\nn}{\nonumber \\ }
\newcommand{\spt}{spacetime}
\newcommand{\ie}{{\em i.e.\/}}
\newcommand{\Mset}{B_{\rm Maxwell}}
\newcommand{\bifset}{{B}}
\newcommand{\crset}{{\cal C}}
\newcommand{\diffset}{{\cal D}}
\newcommand{\R}{{\mathbb R}}
\newcommand{\Z}{{\mathbb Z}}
\newcommand{\MM}{{\cal M}}
\newcommand{\HH}{{\cal H}}
\newcommand{\CS}{{\cal S}}
\newcommand{\UU}{{\cal U}}
\newcommand{\bfem}{\em}
\newcommand{\Cinf}{\mbox{$C^\infty$}}
\newcommand{\paren}[1]{\left({#1}\right)}
\newcommand{\bigparen}[1]{\bigl({#1}\bigr)}

\newcommand{\ug}{unfolding}
\newcommand{\fug}{unfolding}
\newcommand{\ugs}{unfoldings}

\newcommand{\Ugs}{Unfoldings}
\newcommand{\WT}{Whitney {\Cinf} topology}
\newcommand{\Ref}[1]{(\ref{#1})}
\newcommand{\req}{\simeq}
\newcommand{\minimal}{minimum}
\newcommand{\nbhd}{neighbourhood}
\newcommand{\mf}{\mu_F}
\DeclareMathOperator{\rank}{rank}

\begin{document}

\title{%
Topological classification of black Hole: Generic Maxwell set and crease set of horizon}
\author{%
  Masaru Siino\footnote{E-mail:msiino@th.phys.titech.ac.jp} and Tatsuhiko Koike\footnote{E-mail:koike@phys.keio.ac.jp}
}
\address{%
  Department of Physics, Tokyo Institute of Technology, \\
  Oh-Okayama 1-12-1, Megro-ku, Tokyo 152-8550, Japan\\
  Department of Physics, Keio university, \\
  Hiyoshi 3-14-1, Kohoku-ku, Yokohama 223-8522, Japan
}

\begin{abstract}
{}The crease set of an event horizon or a Cauchy horizon
is an important object 
which
determines qualitative properties of the horizon. 
In particular, 
it determines the possible 
topologies of the spatial sections of the horizon.
By Fermat's principle in geometric optics, we relate 
the crease set and the Maxwell set of a smooth function
in the context of singularity theory. 
We thereby give a classification of generic topological 
structure of the Maxwell sets and the generic topologies of the
spatial section of the horizon. 
\end{abstract}
\section{Introduction}
\label{intro}
Topology is one of the fundamental qualitative properties 
of a black hole. 
It was investigated by many authors and now it
is known that, under some reasonable conditions such as asymptotic
flatness and the weak energy condition, 
each component of the black hole region is topologically trivial,
{\ie}, simply connected~\cite{OW}.
On the other hand, there were numerical simulations which suggest
non-trivial topologies of the horizons~\cite{ONW}. 
There has been some confusion, but it is now well understood that 
even though the black hole region in the {\spt} is simply connected, 
there are many possible topologies of spatial sections. 
In particular, 
one of the authors~\cite{MS1} showed how topology of the 
spatial sections of a black hole is related to 
the endpoint set, or similarly, the {\em crease set}, of the event 
horizon. 
Therefore the  crease set of an event horizon,
or of a Cauchy horizon, 
is an important object 
which is independent from the choice of time slices and
which determines qualitative properties of the horizon.
Thus, to restrict the physically possible topologies 
of the black hole it is important to restrict the possible structure 
of the crease set. 

In this paper we classify the possible topological structure of 
the crease set. 
``Possible'' structures are important because 
they are the only ones that actually appear in the real world. 
There are at least two ways to define the ``possible'' structure:
by {\em stability\/} and by {\em genericity}.
In both approaches we consider {\spt}s with different metrics.  
A stable structure is the one that is invariant against small change
of the metric. 
A generic structure is, intuitively, the one that we find 
when he randomly pick up a {\spt}. 
Physically, the small change of the metric may be interpreted 
as follows.
First, in theoretical treatment of the world, we always assume 
some simple evolution equation and equation of state for matter. 
The precise equations may never be known. 
Second, when we are to determine the {\spt} by observations, 
we can never collect perfectly accurate data of field strengths 
or matter density at all points of the {\spt}. 
There must be errors and limits.
Third, if we consider quantum mechanics, the fields and 
the metric can actually fluctuate at small scales. 
We shall give the precise mathematical definitions later. 
One interesting thing is that the both approaches lead the same
conclusion (Theorem~\ref{th:th2}), namely, stability and genericity 
are equivalent. 

Our stability/genericity approach is complementary to the analysis of  
exact solutions such as the Schwarzschild and Kerr {\spt}s
because 
the exact solutions are usually obtained by assuming high symmetry 
and many of its properties, especially topological ones, 
are not stable against perturbations~\cite{MS2}. 
One approach and the results will serve as guiding principles
in the study of  
the black hole {\spt} in mathematical relativity or in astrophysics.   



The studies in the present paper are  common for event
horizons and Cauchy horizons, except for their direction of time. 
Therefore we only consider event horizons in the rest
of the present paper. 
The same results hold for Cauchy horizons.

In Sect.~\ref{fermat}, we define Fermat's potential 
in a general nonstationary {\spt} representing a gravitational
collapse 
which is essential for the crease set and its classification. 
In Sect.~\ref{maxwell}, we give the precise definition of 
the Maxwell set which corresponds to the crease set of the event
horizon. We also define stability and genericity of the Maxwell set, 
and introduce concepts necessary for our investigation. 
We show in Sect.~\ref{class} the equivalence of stability and
genericity and obtain a list of stable Fermat potentials. 
We give in Sect.~\ref{class-max} 
the classification of the stable Maxwell sets. 
In Sect.~\ref{other-dim},  
we discuss the cases of {\spt}s of dimension other than four. 
Sect.~\ref{conc} is for conclusion and discussions. 

For the terms and notations about causal structure of a {\spt}, 
see, e.g., Ref.~\cite{HE}.

\section{Fermat's principle and crease set}
\label{fermat}
An event horizon is generated by null geodesics. 
A future {\eh} $\HH$ 
cannot have future endpoints, but 
can have past endpoints if it is not eternal. 
As is pointed out in \cite{MS1}, 
the {\em endpoint set}\ $\EPS$ of a horizon 
is an arc-wise connected acausal set. 
Points $u\in\EPS$ are classified by 
the {\em multiplicity\/} $m(u)$ of $u$, 
the number of the generators emanating from $u$:
\begin{eqnarray}
  \crset&:=&\{u\in\EPS\;|\;m(u)>1\},\nn
  \diffset&:=&\{u\in\EPS\;|\;m(u)=1\}.
\end{eqnarray}
The set $\crset$ is called the {\em crease set} of the horizon. 
The crease set contains the interior of the endpoint set, 
{\ie},
the closure of $\crset$ contains $\EPS$~\cite{BK}. 
The crease set $\crset$ equals the set of points of $\EPS$ on which the 
horizon is not differentiable, {\ie}, 
the horizon is differentiable at $u\in\EPS$ if and only if
$u\in\diffset$~\cite{BK,Chr98CQG}.

The horizon $\HH$ is the envelope of the light cone starting from the
crease set $\crset$ 
which is an arc-wise connected acausal subset of $\HH$. 
In particular, if the spatial section of the horizon 
is a topological sphere at late times, 
the topology of the spatial section of the horizon can be nontrivial
only at the crease set and the topology is completely determined by
the time slicing of the crease  set.
This is studied in Ref.~\cite{MS1} in detail. 
In particular, when the crease set
is a single point, all the possible spatial section of the horizon is
a topological sphere. On the contrary, when the crease set 
has a disk-like structure, 
the horizon can have toroidal or higher-genus spatial
sections. One would see the coalesce of horizons if 
the crease set has a line-like structure~\cite{ONW}. 
Therefore by classifying the structure of the crease set, we will know
all possible topologies of the horizons. 
Here we do not assume that the spatial section of the horizon in the
future is a sphere.

The crease set can be 
determined by Fermat's principle in a simple stationary {\spt}. 
In a non-stationary {\spt}, 
we can extend Fermat's principle 
and find a variational principle about light paths, 
imposing some appropriate causality condition such as 
global hyperbolicity. 
Here we show an example of the construction of the Fermat potential.

\begin{figure}
\centering
\resizebox{0.75\textwidth}{!}{%
  \includegraphics{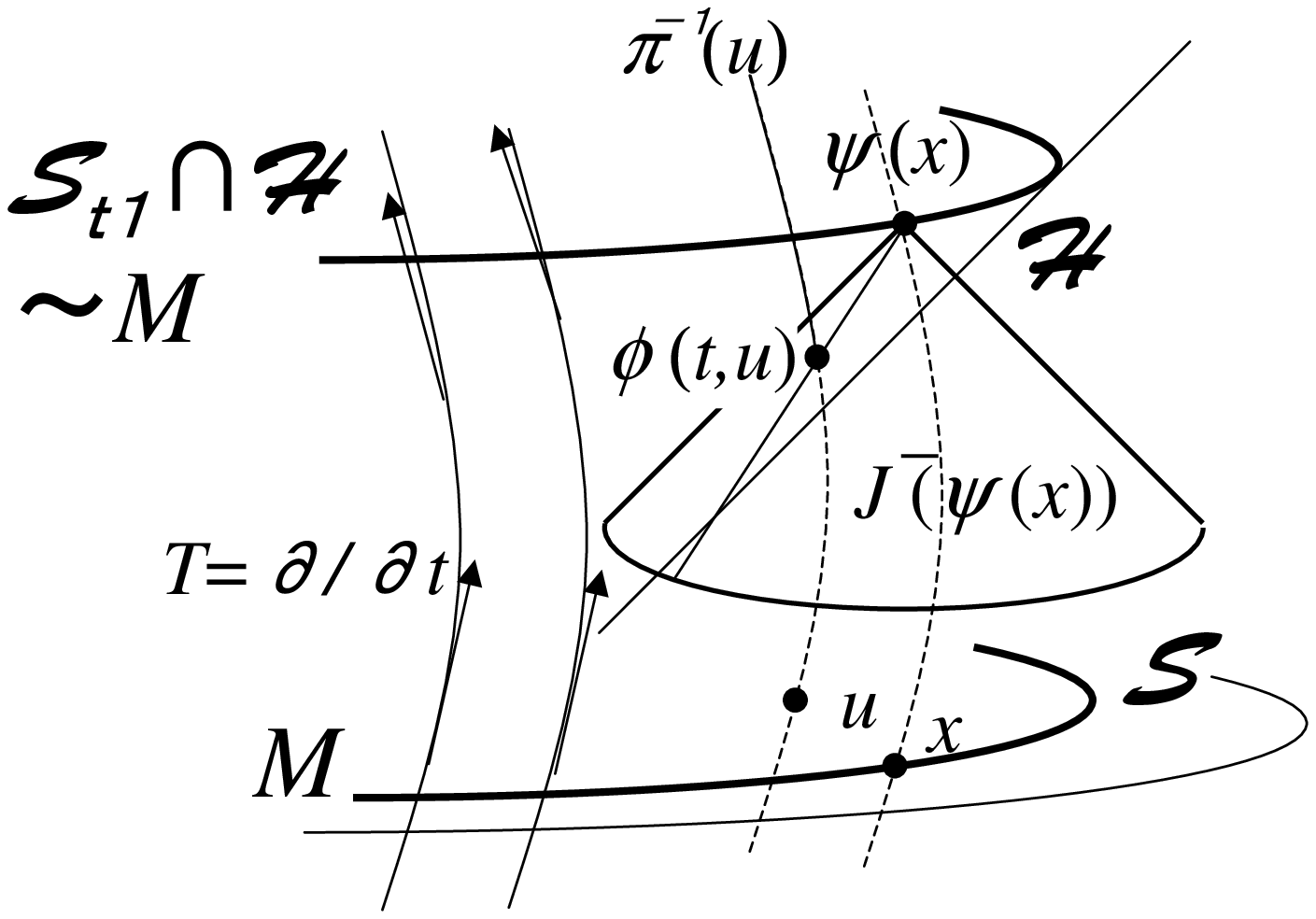}
}
\caption{An example of construction of a Fermat potential. 
  The projection $\pi$ from the {\spt} $\MM$ to a Cauchy surface $\CS$  
  is defined by a timelike vector field 
  $T=\partial/\partial t$. 
  A spatial section of the horizon is give by 
  $\HH\cap \CS_{t_1}=\psi(M)$, where $M$ is a compact subset of $\CS$
  and $\psi=(\pi|_{\HH})^{-1}$. 
  The Fermat potential $F(x,u)$ is defined by 
  minus the supremum of $t$ such that there is a causal curve 
  from $\phi(t,u)$ to $\psi(x)$. 
  Let $x$ is a minimum point of $F(\bullet,u)$ over $M$. 
  Then the null geodesic through $\psi(x)$ which intersects 
  $\phi(t,u)$ gives a null generator of the horizon.}
\label{fig:fermat}       
\end{figure}

Let us assume that the {\spt} $\MM$ is smooth and is 
globally hyperbolic from 
a smooth Cauchy surface $\CS$ which is diffeomorphic to $\R^3$. 
Furthermore, we consider a {\spt} of gravitational collapse, 
namely, we assume that the event horizon $\HH$ is in the future of
$\CS$. 
By global hyperbolicity, there are
always an appropriate smooth global time coordinate $t:\MM\to\R$ and a 
timelike vector field $T$ such that $dt(T)=1$. 
The {\spt} $\MM$ is foliated by Cauchy surfaces 
${\CS}_t=\{q\in\MM|t(q)=t\}$. 
The vector field $T=\partial/\partial t$ 
defines a smooth projection $\pi$ from $\MM$ into the
$\CS={\CS}_{t_0}$ (see
Fig.\ref{fig:fermat}); 
\begin{align}
  &\pi:{\cal M}\to {\cal S},\nn
  &  \pi^{-1}(q)=\{ \gamma(t) |
  \frac{\partial \gamma}{\partial t}=T,
  \gamma(t_0)=q, t\in\R
  \}. 
\end{align}
Conversely, there is a diffeomorphism
\begin{align}
  &\phi:\R\times\CS\to\MM, \nn
  &t(\phi(t,u))=t, \quad \pi(\phi(t,u))=u. 
\end{align}
Because $\HH$ is achronal, the restriction of $\pi$ on $\HH$ 
is injective and has an inverse, which we denote by $\psi$:
\begin{align}
  &\psi:\pi(\HH)\to\HH, \nn
  &\psi(u)\in\HH,\quad \pi(\psi(u))=u. 
\end{align}
The map $\psi$ is Lipschitz. 

We take some (sufficiently large) $t=t_1$ and assume that 
$\CS_{t_1}\cap\HH$ is diffeomorphic to a compact manifold $M$. 
We consider $M$ as a fixed submanifold embedded in $\CS$ so that 
$\CS_{t_1}\cap\HH=\psi(M)$. 
Consider a {\nbhd} $\UU$ of $\pi(\HH\cap J^-(\CS_{t_1}))$ in $\CS$.
For $x\in M$ and $u\in \UU$ we define Fermat's
potential as follows:
\begin{align}
  F(x,u):=-\sup \{t\in\R|\phi(t,u)\in J^-(\psi(x))\}.
  \label{eq-fermat}
\end{align}
The minimum points of $F$ corresponds to the generator of $\HH$
through $x$. 
In particular, when the {\spt} is static, the Fermat potential 
is the spatial geodesic distance, namely, 
\begin{equation}
F(x,u)
=\int_{x}^{u}\sqrt{\gamma_{ij}dx_idx_j}\,+\text{const.}
\label{eq-dist}
\end{equation}
The projection  $\pi$ is generated by the timelike
killing vector and 
$\gamma_{ij}$ is an induced metric of the hypersurface ${\cal S}$
orthogonal to the timelike killing vector. 
Our definition \Ref{eq-fermat} is the generalization of this 
geodesic distance function \Ref{eq-dist} 
to the non-static {\spt}.

From \Ref{eq-fermat}, the crease set $\crset$ is given by 
\begin{align}
  \crset=\psi(\Mset(F)),
\end{align}
where $\Mset(F)$ is the Maxwell set of $F$ where $F$ has two or more
minimum points. We shall give a precise mathematical definition
and a framework to study its properties later. 

Let us consider a small change of the metric on $\MM$. 
To be precise, we may define the small change 
by a $\Cinf$ Whitney topology on the metric tensor field on $\MM$. 
Sufficiently small change of the metric leaves the vector field $T$ 
timelike and the Cauchy surfaces $\CS_t$ spacelike. 
The horizon near $\CS_{t_1}\cap \HH$ changes only slightly in the
{\spt}  $\MM$~\cite{KS}. 
This causes a small change of the diffeomorphism $\psi$ 
from $M$ to $\psi(M)\in\HH$ but the topology of $M$ does not change. 
Also, the null geodesic system hence $J^-$ changes slightly. 
These cause the Fermat potential $F$ to be deformed slightly. 
In the sequel
we shall study the stability and genericity of $F$ 
and $\Mset(F)$ against this deformation of $F$.

In the formulation using such a potential function, actually a state
space becomes 
infinite dimensional manifold, since we should not consider only the
endpoints but also 
the path connecting them. However, if one can restrict the set of path into
a finite dimensional family, the state space reduces to finite
dimensional one~\cite{PS}.
Indeed, in our case, two different points on event horizon cannot be connected
by more than one generators by the fact that the generators of event horizon
cannot have future endpoint. 

On the other hand, in special cases
one can take no spatial section on which there is no singular point of
the horizon~\cite{Kossowski}. 
Of course, such a difficulty is resolved in  other formulations using
``Lagrange manifold'' and the result will not change. 
We are not concerned with this any more in the following.

The goal of our present study is to classify  all generally possible
structure about the singularities of the Fermat potential determining
the horizon. 
As we shall see, the generic structure will be given by studying
singularities of stable Fermat 
 potentials concretely. The genericity or stability is defined as the
 property under small perturbations in a sense of {\WT}. 
 On the other hand, the stable Fermat potential
is equivalent to the universal unfolding (parameter deformation family
of a function) with the same number of deformation parameters. Then
our main task is to give a classification of the universal
 unfoldings.

The above is a usual procedure when one discuss the bifurcation
structure, 
so-called caustics, of a system. 
However, our main object here is not 
the bifurcation set but the Maxwell set (the difference will become
clear later). The problem is not purely local but is 
rather semi-local. The
definition 
of Maxwell set is local in control (parameter) space $U$ but
is non-local in state (variable) space $M$. 
To treat this we introduce function multigerms.
We will classify the universal multiunfoldings 
and the Maxwell sets. 

Finally, we note that 
the catastrophe changes
when some symmetry is present. 
An example is
the toroidal event horizon studied in Refs. \cite{ONW}. 
This is not a universal unfolding but an unfolding with infinite
codimension.

\section{The Maxwell set: Stability and genericity}
\label{maxwell}
Now we give the mathematical framework to study the Maxwell set.
The Maxwell set of a function unfolding is the set of all values of
the parameters 
for which 
the minimum is attained either at a non-Morse
critical point or at two or more critical points.
In the following we sometimes make $x$ as a
representative of the state variables and $u$ the control variables.
Of course, $F$ is not a global function on ${\R}^n\times {\R}^r$
rather a function 
on manifold $M\times U$, where $M$ is two-dimensional compact manifold
and $U$ is an open subset of ${\R}^3$. 
We will mainly deal with function germs 
instead of functions, since we 
are only concerned with local properties of a function on some
{\nbhd} of $u_0\in U$,  
To treat this, we introduce a germ of map which is an equivalence
class of maps 
at a {\nbhd} $W$. 
A function $F:M\times U\to {\R}$ can be considered as 
a family of functions $f_u:M\to\R$ with $f_u=F(\bullet,u)$.
\begin{definition}[Unfolding]
A function $F:M\times U\to {\R}$ is 
called an {\bfem unfolding\/} 
of a function $f_{u_0}$ {\em at\/} $u_0$. 
\end{definition}
\begin{definition}[Maxwell set]
For a function unfolding $F:M\times U\to {\R}$ on a compact
manifold $M$, 
the {\bfem Maxwell set} ${\Mset}(F)$ of $F$ is 
a subset of $U$ given by
\begin{eqnarray}
{\Mset}(F)&:=&\{u\in U| \text{$f_u$ has two or more }
\text{global {\minimal}  
  points}\}. 
\end{eqnarray}
\end{definition}

In the investigation of the Maxwell set
we mainly focus on its local structure because 
the global structure is obtained by the combinations of local ones.
Below we extensively use the notion of the germs of objects 
which provides the best way to characterize their local structure. 
Let $M$, $N$ be \Cinf-manifolds. 
We denote the set of $\Cinf$-maps from $M$ to $N$ by
$\Cinf(M,N)$. 
\begin{definition}[Map germ]
Maps $f,g\in\Cinf(M, N)$ 
are {\bfem equivalent} 
at $x_0$ if there is a {\nbhd}
$W$ of $x_0$ such that $f|_{W}=g|_{W}$. 
A {\bfem map germ\/} $f$ at $x_0$, $[f]_{x_0}$,  
is the equivalence class 
of $f$.
It is also denoted by 
$f:(M,x_0)\to N$ or 
$f:(M,x_0)\to(N,f(x_0))$. 
\end{definition}
Examples of map germs include function germs and diffeomorphism
germs.

\begin{definition}[Set germ]
Subsets $X$, $Y$ of 
$M$ 
are {\bfem equivalent at\/} $x_0$ if 
$X\cap W=Y\cap W$ for each {\nbhd} $W$ of $x_0$. 
A {\bfem set germ\/} $(X,x_0)$ of $X$ at $x_0$ is the equivalence class 
of $X$. 
Set germs $(X,x_0)$ and $(Y,y_0)$ are {\bfem diffeomorphic\/}, 
$(X,x_0)\simeq(Y,y_0)$,  
if there is a diffeomorphism germ $\phi:(M,x_0)\to M$ 
such that 
$(\phi(X),\phi(x_0))=(Y,y_0)$.
\end{definition}

\begin{definition}[Unfolding germ]
Function unfoldings $F$ and $G$ are {\bfem equivalent\/} 
at $u_0$ 
if 
there is
a {\nbhd} $W$ of $u_0$ in $U$ such that 
$F|_{M\times W}=G|_{M\times W}$.
An {\bfem unfolding germ\/} 
$
F:(M\times U,M\times \{u_0\})\to {\R}
$, or $[F]_{u_0}$, 
of $f_{u_0}$ at $u_0$ is 
the equivalence class of $F$ 
defined by this equivalence relation. 
\end{definition}
We usually call unfolding germs defined above, 
which are germs with respect
to $U$ only, simply as unfoldings. 
Later we will define unfolding germs with respect to both $M$ and
$U$, which we will call unfolding germs. 






Below, we will determine the topological structure of $\Cinf(M,N)$
where all the maps that we treat are included. We define a topology of
$\Cinf(M,N)$ by the $r$-jet space $J^r(M,N)$ below.  

\begin{definition}[Jet space]
Let $f\in \Cinf(M,N)$. 
The {\bfem $r$-jet} $j^rf(x_0)$ of $f$ 
at $x_0$ is the equivalence class 
of $f$ in $\Cinf(M,N)$
where 
two maps are equivalent if
all of their $s$-th partial derivatives with $1\le s\le r$,
in some coordinate systems of $M$ and $N$, coincide. 
The {\bfem $r$-jet space\/} of $\Cinf(M,N)$
is defined by 
\begin{equation}
J^r(M,N):=\{j^rf(x_0)|f\in \Cinf(M,N)\}. 
\end{equation}
\end{definition}
The space of $r$-jets at a point is 
an $n\vtr{m+r}r$-dimensional manifold, 
where $m=\dim M$ and $n=\dim N$. 


Now we endow the space $\Cinf(M,N)$ with the {\WT}. 
\begin{definition}[{\WT}]
For an open subset $O$ of $J^r(M,N)$, let
\begin{equation}
W^r(O):=\{f\in \Cinf(M,N)|j^rf(M)\subset O\}.
\end{equation}
The {\bfem Whitney \Cinf topology} on $\Cinf(M,N)$ 
is the topology whose basis is 
\begin{eqnarray}
  \bigcup_{r=0}^\infty 
  \{W^r(O)|
  \text{$O$ is an open subset of $J^r(M,N)$}\}.
\end{eqnarray}
\end{definition} 
Hereafter we treat that $\Cinf(M,N)$ as a topological space with 
the {\WT}. 
Now we can define stability of the Maxwell set 
using this topology.
\begin{definition}[Stable Maxwell set germ] 
An {\ug}, 
\begin{equation}
F:(M\times U,M\times \{u_0\})\to {\R}, 
\end{equation}
is {\bfem stable with respect to the Maxwell set\/} 
if for each {\nbhd} $W$ of
$u_0$ there exists a {\nbhd} ${\cal U}$ of $F$ in
$\Cinf(M\times U,{\R})$ such that for each $G\in {\cal U}$ there
exists $v_0\in W$ 
such that $(\Mset(F),u_0)\simeq({\Mset}(G),v_0)$.
We call $({\Mset}(F),u_0)$ a {\bfem
  stable Maxwell set germ}. 
\end{definition}
In this sense, our aim is to classify the stable Maxwell set germs 
of the Fermat potential $F$. 

\label{genericity}

Since we have defined the topology of $\Cinf(M,N)$, now we can
formulate the genericity of a class of smooth maps. 
A subset of a topological space $X$ is 
nowhere dense if its closure has no interior.
A subset of $X$ is {\em residual\/} if 
its complement is a countable union of nowhere dense sets. 
The space $X$ is a {\em Baire space\/} if every residual set is dense. 
\begin{definition}[Genericity]
A {property} $P$ of $f\in \Cinf(M,N)$ is {\bfem generic\/} if the set 
\begin{equation}
{\cal A}_P:=\{f\in \Cinf(M,N)|\text{$P(f)$ is true}\}
\end{equation}
is residual in $\Cinf(M,N)$.
\end{definition}

When $P$ is generic, ${\cal A}_P$ is dense in $\Cinf(M,N)$ and 
any $g\in \Cinf(M,N)$ 
is approximated by a map $f$ satisfying $P$.
$f\in{\cal A}_P$, 
Furthermore, ${\cal A}^c_P$, the set on which the negation of $P$ holds,  
is not generic. 
This is because of the following~(e.g. \cite{BS}).:
\begin{theorem}
  $\Cinf(M,N)$ with the {\WT} is a Baire space.
\end{theorem}

To prove a generic property,  transversality theorems are fundamental.
In particular, we make use of the Multitransversality Theorem by
Mather below. 

\begin{definition}[Transversality]
Let $f\in\Cinf(M,N)$ and let $Q$ be a submanifold of $N$. 
The map $f$ is {\bfem transversal\/} to $Q$ at $x$ if 
either of the following holds: 

(1) $f_*(T_xM)\oplus
T_{f(x)}Q=T_{f(x)}N$, 

(2) $f(x)\notin Q.$\\
The map $f$ is transversal to $Q$ if it is transversal to $Q$ at every $x\in
Q$. 
\end{definition}

\begin{theorem}[Multitransversality Theorem by Mather~\cite{MA}]
Let $M$, $N$ be $\Cinf$-manifolds and 
let $Q_1, Q_2,...$ be a countable family of submanifolds of
${}_kJ^r(M,N)$.  
Then the set 
\begin{equation}
T:=\{f\in \Cinf(M,N)|\text{${}_kj^rf$ is transversal to $Q_1,Q_2,...$}\}
\end{equation}
is residual in $\Cinf(M,N)$. 
\label{th:mth}
\end{theorem}
This theorem is fundamental in our discussion. 
It states that 
transversality to a countable family of submanifolds is a generic property. 
Therefore the properties deduced from this theorem will also be generic.
In the rest of the paper, we shall 
show that any stability and genericity of 
a Maxwell set is equivalent in a certain sense 
(Theorem~\ref{th:th2})~\footnote{%
This is true when $(\dim M,\dim N)$ are nice
  dimensions~\cite{ND}. For example, if the $\dim U>5$,
  stability does not imply genericty. 
  In our case, however, we have $N=\R$ and 
  it is known that $(\dim M,\dim R)=(\bullet,1)$ 
  are always nice dimensions.}
and carry out a topological classification of stable Maxwell sets. 



\section{Classification of the Fermat potentials}
\label{class}
Let us give the classification of stable Maxwell sets. 
The definition of the Maxwell set requires
the global information of the function unfolding $F$. 
A simple but crucial observation is, however, that 
to determine the local structure of the Maxwell set, \ie, 
the Maxwell set germs, 
we only need the {\em local information of $F$ around its 
global {\minimal}  points\/} $p_1,...,p_k\in M$. 
We first generalize the notion of germs to that of multigerms.  
Let $M^{(k)}$ be a $k$-tuple of distinct points of $M$, \ie,
\begin{equation}
M^{(k)}:=\{(x_1,...,x_k)\in M^k|x_i\neq x_j\text{ for }i\neq j\}. 
\end{equation}
\begin{definition}[Multigerm]
  Let $(x_1,...,x_k)\in M^{(k)}$. 
  A {\em $k$-fold map germ\/} 
  $f:(M,(x_1,...,x_k))\to N$,
  or 
  $[f]_{x_1,...,x_k}$,
  is the equivalence class of $f\in\Cinf(M,N)$,  
  where two maps are equivalent if they coincide 
  on some open subset of $M$ which contains 
  $x_1,...,x_k$. 
  A {\em $k$-fold unfolding germ\/} 
  $F:(M\times U,((x_1,...,x_k),u_0))\to N$, 
  or $[F]_{(x_1,...,x_k),u_0}$, 
  is the equivalence class of 
  $F\in\Cinf(M,\R)$ 
  where two functions are equivalent if they coincide 
  on some open subset of $M\times W$ 
  which contains $(x_1,u_0),...,(x_k,u_0)$. 
  A $k$-fold germ is also called as a {\em multigerm}. 
\end{definition}
A multigerm can be considered as $k$-tuple of simple germs. 
For example, A map multigerm $[F]_{(x_1,...,x_k),u_0}$ 
can be considered as $k$-tuple of function germs 
$([f_1]_{x_1},...,[f_k]_{x_k})$. 

\begin{definition}[Right equivalence]
Function germs 
$f:(M,x)\to {\R}$ and 
$g:(N,y)\to {\R}$ 
are {\bfem right equivalent\/}, 
$[f]_{x}\req [g]_{y}$, 
if there exist a diffeomorphism germ 
$\phi:(M,x)\to(N,y)$
and $a\in\R$ 
such that $f=g\circ \phi+a$ holds as an equality of 
function germs at $x$. 
\end{definition}
\begin{definition}[Right equivalence at the {\minimal} points]
{\Ugs} $F:(M\times U,M\times\{u_0\})\to {\R}$ and 
$G:(N\times V,N\times\{v_0\})\to {\R}$ 
are {\bfem right equivalent at the  {\minimal} 
points\/}
if the following conditions hold: 

(1) The functions $f_{u_0}=F(\bullet,u_0)$ and 
$g_{v_0}=G(\bullet,v_0)$ have
the same number of global {\minimal}  points, 
$p_1,...,p_k$ and $q_1,...,q_k$, respectively. 

(2) 
There exist 
  a diffeomorphism multigerm
  $
    \phi:(M \times U,((p_1,...,p_k),u_0))
  \to
  (N \times V,((q_1,...,p_k),v_0))
  $, 
a diffeomorphism germ 
  $
    \psi:(U,u_0)\to(V,v_0)
  $, 
  and
  a function germ
  $\alpha:(U,u_0)\to\R$
  such that
\begin{equation}
  \label{eq-sta}
  F(x,u)=G(\phi(x,u),\psi(u))+\alpha(u)
\end{equation}
holds with both sides being 
function multigerms at $((p_1,...,p_k),u_0)$. 
\end{definition}




The Maxwell set germ of an {\ug} 
is determined only by the 
unfolding multigerm at the  {\minimal}  points: 
\begin{proposition}
If {\ugs} 
$F:(M\times U,M\times\{u_0\})\to\R$ 
and
$G:(M\times V,M\times\{v_0\})\to\R$ 
are right equivalent at the {\minimal} points, 
then their Maxwell set germs are diffeomorphic. 
\label{col:1}
\end{proposition}
\begin{proof}
  Follows directly from 
  the definitions of right equivalence and 
  of Maxwell set germs. 
\end{proof}

Now we have completed the preparation. 
We will investigate stable structure of the function unfolding and its
Maxwell set. 

\begin{definition}[Stability at the minimum points]
An {\ug} 
$F:(M\times U, M\times \{u_0\})\to{\R}$
is {\bfem stable at the  {\minimal}  points} 
if
for each {\nbhd} $W$ of $u_0$ 
there exists a {\nbhd} ${\cal U}$ of $F$ in 
$\Cinf(M\times U,{\R}$) 
such that for each $G\in{\cal U}$ there exists $v_0\in W$ 
such that 
$[F]_{u_0}\req [G]_{v_0}$. 
\end{definition}
From Proposition~\ref{col:1}, we immediately 
have the following proposition. 
\begin{proposition}
If an {\ug} $F:(M\times U, M\times \{u_0\})\to{\R}$
is stable at the  {\minimal}  points then 
$({\Mset}(F),u_0)$ is a stable Maxwell set germ. 
\end{proposition}

To discuss the stability, we study the orbit of diffeomorphism on some
standard functions in $J^r(M,N)$ and its transversality. 
We will stratify $J^r(M,N)$, {\ie}, 
decompose the $J^r(M,N)$ into the union of submanifolds 
(strata)~\cite{STF}.  

\begin{definition}[Strata]
\begin{align}
A_0&:=\{j^rf(p)\in J^r(M,{\R})\,|\, 
\text{$f$ is regular at $p$}\},\\
A_k&:=\{j^rf(p)\in J^r(M,{\R})\,|\,
[f]_p\req
[\pm x_1^{k+1}\pm x_2^2]_0\},\\ 
D_4&:=\{j^rf(p)\in J^r(M,{\R})\,|\,
[f]_p\req
[  \pm x_1^3\pm x_1x_2^2]_0\},\\
E_5&:=\{j^rf(p)\in J^r(M,{\R})\,|\,
[f]_p\req
[\pm x_1^3\pm x_2^3]_0\}. 
\end{align}
\end{definition}
The following is well discussed~\cite{STF}:
\begin{lemma}
  (1) $A_0$ is an open subset of $J^r(M,{\R})$ 
  hence is a submanifold of 
  codimension 0. 

(2) $A_k$ is a submanifold of $ J^r(M,{\R})$ of codimension $k+1$.

(3) $D_4$ and $E_5$ are submanifolds of $J^r(M,{\R})$ of 
  codimension 5.

(4)  $\Sigma :=J^r(M,{\R}) -\bigcup_{k=0}^4A_k - D_4-E_5$ 
  is the union of a finite number of submanifolds 
  of codimension $6$ or greater:
\begin{equation}
\Sigma=W_1\cup ...\cup W_s. 
\end{equation}
\end{lemma}


\begin{definition}[Natural stratification]
The {\bfem natural stratification\/} of $J^r(M, {\R})$, where $r>4$, 
is the one given by
\begin{align}
{\cal S}(J^r(M,{\R}))
:=\{A_0,A_1,...,A_4,D_4,E_5,W_1,...,W_s \}.
\end{align}
\end{definition}

We also extend the concept of jet space to multijet space.
\begin{definition}[Multijet space]
The {\em $k$-fold  $r$-jet\/},  or 
simply, {\bfem $r$-multijet\/}, 
${}_kj^rf$ of $f:M\to\R$ at $(x_1,...,x_k)\in M^{(k)}$ is
\begin{align}
  {}_kj^rf(x_1,...,x_k):=\paren{j^rf(x_1),...,j^rf(x_k)}. 
\end{align}
The {\bfem $r$-multijet space} 
${}_kJ^r(M,N)$ is given by
\begin{align}
{}_kJ^r(M,N)&:=\{{}_kj^rf(x_1,...,x_k)\in
(J^r(M,N))^k|
\nn
&\quad (x_1,...,x_k)\in M^{(k)}\}. 
\end{align}
The map
$
  {}_kj^rf:M^{(k)}\to {}_kJ^r(M,N)
$
is called an {\em $r$-multijet section\/}. 
\end{definition}

To classify event horizons in a four-dimensional {\spt}, 
we assume that $M$ is a two-dimensional compact manifold and
$U$ is diffeomorphic to $\R^3$. 
Then, because $\dim (M\times U)=5$, 
it is sufficient to consider the 5-fold 5-jet, ${}_5J^5(M,{\R})$, 
Let us define a natural stratification of the multijet
space ${}_kJ^5(M,{\R})$. 

\begin{definition}[Natural stratification of ${}_kJ^5(M,{\R})$]
The {\bfem natural stratification\/} of ${}_kJ^5(M, {\R})$ is the one
given by 
\begin{align}
{\cal S}({}_kJ^5(M,{\R})):=&\{\Delta_k\cap X_1\times ...\times X_k
|
X_1,...,X_k \in {\cal S}(J^5(M,{\R}))\}, 
\end{align}
where 
\begin{align}
\Delta_k := &\bigl\{ (j^5f_1(p_1),...,j^5f_k(p_k)) 
\in
{}_kJ^5(M,{\R}) \big|
f_1(p_1) = ... =f_k(p_k)\bigr\}. 
\end{align}
\end{definition}

The following lemma 
is easily shown. 
\begin{lemma}
Let $M$, $N$ be $\Cinf$-manifolds and 
let $f:M\to N$ be a $\Cinf$-map. 
Let $S$ be a 
a $\Cinf$-submanifold of codimension $c$. 
Then for each $x\in M$ there exists 
a {\nbhd} $W$ of $f(p)$ in $N$ and 
a $\Cinf$-map $g:W\to {\R}^c$
such that  
\begin{eqnarray}
&&\rank dg_{\xi}=c, 
\quad
S\cap W=g^{-1}(0), \quad 0\in\R^c, 
\label{eq:lm42}
\end{eqnarray}
where $dg_{\xi}:T_\xi\to T_{g(\xi)}\R^c$ 
is the differential map of $g$ at $\xi$. 
\label{lm:42}
Moreover, the following two conditions are equivalent:

(1) $f$ is transversal to $S$ at $x$.

(2) $\rank d(g\circ f)_x=k$.
\end{lemma}


\begin{theorem}
Let $F: (M\times U, M\times \{u_0\})\to {\R}$ be an {\fug} 
and 
let $p_1,...,p_k$ be 
the {\minimal}  points of $f_{u_0}=F(\bullet,u_0)$. 
The unfolding $F$ is stable at the {\minimal}  points if
and only if the multijet section 
\begin{equation}
{}_kj^5F: M^{(k)}\times U\to {}_kJ^5(M,{\R})
\end{equation}
is transversal to the natural stratification 
${\cal S}({}_kJ^5(M,{\R}))$ at 
$((p_1,...,p_k),u_0)\in (\R^2)^k\times U$. 
Furthermore, such $F$ falls into one of the following four cases. 
(Below $0_i\in \R^2$ are the origins of local coordinate systems 
$(x_i,y_i)$). 

(1) 
$k=1$ and $F$ is right equivalent at the minimum points 
to either 
\begin{align}
A_1(x,y,u)&=x^2+y^2
\text{ at $(0,0)\in{\R}^2\times{\R}^3$}, 
\label{eq-A1}
\end{align}
or
\begin{align}
A_3(x,y,u)&=x^4+u_2x^2+u_1x+y^2 
\text{ at $(0,0)\in{\R}^2\times{\R}^3$}. 
\label{eq-A3}
\end{align}

(2) 
$k=2$ and $F$ is right equivalent at the minimum points 
to either 
\begin{align}
(A_1,A_1)=&(x_1^2+y_1^2+u_1,x_2^2+y_2^2) 
\text{at $((0_1,0_2),0)\in (\R^2)^{(2)}\times\R^3$},
\end{align}
or
\begin{align}
(A_3,A_1)=&(x_1^4+u_2x_1^2+u_1x_1+y_1^2+u_3,x_2^2+y_2^2)
\nn
&\text{at $((0_1,0_2),0)\in (\R^2)^{(2)}\times\R^3$}.
\end{align}

(3) 
$k=3$ and $F$ is right equivalent at the minimum points to 
\begin{align}
  (A_1,A_1,A_1)=&(x_1^2+y_1^2+u_1,x_2^2+y_2^2+u_2,x_3^2+y_3^2)
  \nn
  &\text{at $((0_1,0_2,0_3),0)\in (\R^2)^{(3)}\times\R^3$}. 
\end{align}

(4)
$k=4$ and $F$ is right equivalent to 
\begin{align}
&(A_1,A_1,A_1,A_1)\\
=&
(x_1^2+y_1^2+u_1,x_2^2+y_2^2+u_2,x_3^2+y_3^2+u_3,x_4^2+y_4^2)
\nn
&\text{at $((0_1,0_2,0_3,0_4),0)\in (\R^2)^{(4)}\times\R^3$.} 
\end{align}
\label{th:th2}
\end{theorem}
\begin{remark}
  The theorem states that 
  {stability} and genericity are equivalent 
  in the following sense. 
  Stability at the minimum points,
  or, equivalently, being in either one of the four cases above, is
  a generic property of $F$. 
  On the contrary, when we classify unfoldings 
  by right equivalence at the minimum points, 
  the set of all generic unfoldings must include the four cases above. 
  This equivalence exists for any {\spt} dimensions. 
\end{remark}

\begin{proof}

{\em (i) The stability 
  implies the multitransversality.} 

Let $F: (M\times U, M\times \{u_0\})\to {\R}$ 
be stable at the {\minimal}  points. 
Then there exists 
a {\nbhd} ${\cal U}$ of $F$ in $\Cinf(M\times U,\R)$ 
such that for each $G\in {\cal U}$ 
there exists $v_0\in U$ such that 
\begin{equation}
  \label{eq-tmp}
  F(x,u)=G(\phi(x,u),\psi(u))+\alpha(u)
\end{equation}
with both sides being 
function multigerms at $((p_1,...,p_k),u_0)$, 
where $\phi$, $\psi$ and $\alpha$ are as those 
in \Ref{eq-sta}. 
From the Multitransversality Theorem, 
the unfolding 
$G:(M\times U,M\times \{v_0\})\to {\R}$ can be chosen 
so that its multijet section 
${}_kj^5G$ 
is transversal to 
${\cal S}({}_kJ^5(M,{\R}))$.
By the definition of ${\cal S}({}_kJ^5(M,{\R}))$, 
the action of $\phi$ on 
${}_kJ^5(M,{\R})$
preserves the stratification ${\cal S}({}_kJ^5(M,{\R}))$. 
Moreover, the term $\alpha(u)$ in \Ref{eq-tmp} is irrelevant
concerning the transversality. 
Thus, from \Ref{eq-tmp}, 
${}_kj^5F$ is transversal to
${\cal S}({}_kJ^5(M,{\R}))$. 

{\em (ii) The multitransversality implies that 
one of the conditions (1)--(4) holds.} 

Let 
the multijet section 
${}_kj^5F: M^{(k)}\times U\to {}_kJ^5(M,{\R})$
be transversal to 
${\cal S}({}_kJ^5(M,{\R}))$ at 
$((p_1,...,p_k),u_0)$. 

Suppose $k\geq 5$. Then 
the transversality of  
the multijet section ${}_kj^5F$ 
to ${\cal S}({}_kJ^5(M,{\R}))$ at $((p_1,...,p_k),u_0)$ 
would imply
\begin{equation}
{}_kj^5F((p_1,...,p_k),u_0)\notin \Delta_k\cap
A_1\times...\times A_1 . 
\end{equation}
because $\text{codim} \paren{\Delta_k\cap A_1\times...\times
A_1}=k+1>5$ and $\text{dim}(M\times U)=5$. 
For the other strata 
${\Delta_k\cap X_1\times...\times X_k}$ 
with $X_i\ne A_0$, 
the codimension is larger and we would have
${}_kj^5F((p_1,...,p_k),u_0)\notin \Delta_k\cap
X_1\times...\times X_k$.  This would contradict 
with the fact that $p_1,...,p_k$ are minimum points.
Thus the cases with $k\ge5$ cannot occur. 

When $k=1$, 
the statement is 
a direct consequence of Thom's elementary catastrophe theory. 

When $2\le k\le 4$, 
let the multijet section ${}_kj^5F$ be transversal to
${\cal S}({}_kJ^5(M,{\R}))$ at $((p_1,...,p_k),u_0)$, 
where  $p_1,...,p_k$ are the {\minimal}  points of $f_{u_0}$. 
Then the jet section $j^5F$ 
must be transversal to ${\cal S}(J^5(M,{\R}))$
at $(p_i,u_0)$, $i=1,...,k$. 
It follows from the case $k=1$ that 
$j^5F$ must be in either $A_1$ or $A_3$.
Thus by considering 
the codimensions of $A_k$, $D_4$, $E_5$ and $\Sigma$, 
the dimension of $M\times U$ (=5), 
and the fact 
that $p_1,...,p_k$ are {\minimal}  points,
we find that the only possibilities are the following, 
up to addition of a function of $u$:\\
(a) $k=2,3,4$ and 
\begin{align}
F(x_i,y_i,u_1,u_2,u_3)&=x_i^2+y_i^2+\alpha_i(u), \quad(i<k)\nn
F(x_k,y_k,u_1,u_2,u_3)&=x_k^4+y_k^2,
\end{align}
(b) 
$k=2$ and
\begin{align}
F(x_1,y_1,u_1,u_2,u_3)&=x_1^4+y_1^2+u_2x_1^2+u_1x_1+\alpha_1(u),\nn
F(x_2,y_2,u_1,u_2,u_3)&=x_2^2+y_2^2,
\end{align}
where for $i=1,...,k$, 
$(x_i,y_i,u_1,u_2,u_3)$ is 
some local coordinate system in a {\nbhd} of 
$(p_i,u_0)$ in $M\times U$.

In the following we omit detailed calculations and sketch the proof.
In the case (a), one can easily construct 
a map $g:{}_kJ^5(M,\R)\to\R^{3k-1}$ 
such that 
$\Delta_{(p_1,...,p_k)}\cap A_1\times...\times A_1=g^{-1}(0)$,
where $\Delta_{(p_1,...,p_k)}$ is $\Delta_k$ with the minimum points
fixed to $(p_1,...,p_k)$. 
One can verify that the differential map $dg_\xi$ is nondegenerate 
for $\xi\in {}_kJ^5(M,\R)$. 
By
transversality of 
${}_kj^5F$ to 
${\cal S}({}_kJ^5(M,{\R}))$ at 
$((p_1,...,p_k),u_0)$ 
and by Lemma~\ref{lm:42}, 
we find that
$d(g\circ {}_kj^5F)_{((p_1,...,p_k),u_0)}$ is nondegenerate. 
This implies that the Jacobi matrix of the map 
$(u_1,u_2,u_3)\mapsto
(\alpha_1(u),...,\alpha_{k-1}(u))$ 
is nondegenerate. 
This allows a coordinate transformation 
$(u_1,u_2,u_3)\mapsto(v_1,v_2,v_3)$ given by 
\begin{align}
  v_i&=\alpha_i(u), \ i\le k-1,\quad
  v_i=u_i, \ i>k-1.
\end{align}
This gives the normal forms in the theorem.

In the case (b), we construct 
$g:{}_kJ^5(M,\R)\to\R$ 
such that $\Delta_2=g^{-1}(0)$ where $dg_\xi$ 
is nondegenerate. 
By
transversality of 
${}_kj^5F$ to 
$\Delta_2\cap A_3\cap A_1$ at 
$((p_1,p_2),u_0)$ 
and by Lemma~\ref{lm:42}, 
we find that
$d(g\circ {}_kj^5F)_{((p_1,p_2),u_0)}$ is nondegenerate. 
This implies that the following coordinate transformation 
is possible:
\begin{align}
  v_1=u_1,\quad
  v_2=u_2,\quad
  v_3=\alpha_1(u). 
\end{align}
This gives the normal form in the theorem. 

(iii)
{\em Each of conditions (1)--(4) implies the stability at the minimum
  points.}  

Let us consider the case $k=2$, \ie, 
let $F:(M\times U,M\times \{u_0\})\to {\R}$ 
satisfy (2) in the theorem. 
When $F$ is right equivalent to $(A_3,A_1)$, 
the double 5-jet section
$
{}_2j^5F
$
intersects $\Delta_2\cap A_3\times A_1\in
{\cal S}({}_2J^5(M,{\R}))$ 
transversally at $((p_1,p_2),u_0)\in M^{(2)}\times U$. 
Then, for $G:M\times U\to {\R}$
sufficiently close to $F$, there exists $v_0\in U$
close to $u_0$ such that the number of {\minimal}  points of
$g_{v_0}=G(\bullet,v_0)$ is two and that 
$
{}_2j^5G
$
intersects $\Delta_2\cap A_3\times A_1$ 
transversally at $((q_1,q_2),v_0)\in M^{(2)}\times U$, 
where $q_1,q_2$ are the {\minimal}  points of $g_{v_0}$. 
Then 
from (ii) again, 
$G$ is right equivalent at the {\minimal}
points to $(A_3,A_1)$. 
Since both $F$ and $G$ are right equivalent at the
{\minimal}  points to $(A_3,A_1)$, 
$G$ is right equivalent to $F$ at the {\minimal}
points. Therefore $F$ is stable at
the {\minimal}  points. 
For the other cases, the assertion can be shown similarly.
\end{proof}

\begin{remark}
\label{rem:arn}
In the book of Arnold~\cite{AN} there is a statement in general
dimensions without proof: 
{\em 
The Maxwell set of a generic $l\leq 6$-parameter family of functions is 
locally stably diffeomorphic to one of the sets $(A_{\mu_i})$,
 where the $\mu_i$ are odd and $\sum\mu_i\leq l+1$, i.e., either
 to the set with the singularity $A^m_1(m\leq l+1)$ or to one of
 the sets of the following table (the types
of minimum points are indicated):}
\begin{tabular}{c|c|c|c|c|c}
\hline\noalign{\smallskip}
$l\leq$ & 2 & 3 & 4 & 5 & 6 \cr
\noalign{\smallskip}\hline\noalign{\smallskip}
Type    & $A_3$ & $A_1A_3$ & $A^2_1A_3,A_5$ & $A^2_3,A^3_1A_3,A_1A_5$ &
 $A_1A^2_3,A^4_1A_3,A^2_1A_5,A_7$ \cr
\noalign{\smallskip}\hline
\end{tabular}
\\
The notation is according to his book.
$A_1^2$ means $(A_1,A_1)$, etc.
\end{remark}

\section{Classification of Maxwell sets}
\label{class-max}


In this section we reveal the concrete structure of the Maxwell set
for each $F$ appeared in Theorem~\ref{th:th2}. 


\begin{definition}[Minimum function]
The {\em minimum function\/} ${\mf}:U\to\R$ of 
an unfolding $F: M\times U\to {\R}$ 
is given by 
\begin{equation}
  {\mf}(u):=\min \{F(x,u)| x\in M\}.
\end{equation}
\end{definition}

There are two cases when the minimum function ${\mf}$ has a singularity
at $u$:\\
(1) the function $f_{u}$ has
several  {\minimal}  points in $M$, or \\
(2) the number of minimum points changes there. \\
In the case (1) $u$ is a point of  the Maxwell set. 
In the case (2) $u$ is not a point of the Maxwell set but 
corresponds to a point of the endpoint set $\EPS$. 

In this paper, we define the boundary of a Maxwell set 
as follows. 
An {\em interior point\/} $u$ of a Maxwell set $\Mset(F)$ 
is such that there exists a $C^0$-submanifold $V$ of codimension 1 of 
$U$ which contains $q$ and which is entirely contained in $\Mset(F)$. 
The  {\em boundary\/} of $\Mset(F)$ is the complement of the 
interior of $\Mset(F)$ in $\overline{\Mset(F)}$, where 
$\overline{\Mset(F)}$ is the closure of $\Mset(F)$ in $U$.

\subsection{$A_1$}
In this case the number of {\minimal}  point is one. 
Obviously, the Maxwell set ${\Mset}(A_1)$ is empty. 
The minimum function ${\mf}$ does not have singularities. 

\subsection{$(A_1,...,A_1)$}
In this case of $(A_1,...,A_1)$, 
$k$ {\minimal}  points compete, where $k$ can be 
2, 3, or 4. 
When $k=2$, $F$ is right equivalent at the minimum points
to 
\begin{align}
(A_1,A_1)=&(x_1^2+y_1^2+u_1,x_2^2+y_2^2)
\text{ at $((0_1,0_2),0)\in (\R^2)^{(2)}\times\R^3$}. 
\end{align}
The minimum function is given by
\begin{align}
  \mf(u)=\min\{u_1,0\}
\end{align}
This is singular on the plane $u_1=0$. 
The Maxwell set germ is given by
\begin{align}
  (\Mset(F),0)=(\{(u_1,u_2,u_3)\in\R^3|u_1=0\},0). 
\end{align}
This is a surface (germ) through the origin. 
The points of $(A_1,A_1)$ form a surface. 

When $k=3$, $F$ is right equivalent 
at the minimum points 
to
\begin{align}
  (A_1,A_1,A_1)=&(x_1^2+y_1^2+u_1,x_2^2+y_2^2+u_2,x_3^2+y_3^2)
  \nn
  &\text{at $((0_1,0_2,0_3),0)\in (\R^2)^{(3)}\times\R^3$}. 
\end{align}
The minimum function is
\begin{equation}
  {\mf}(u)=\min\{u_1,u_2,0\}.
\end{equation}
The Maxwell set is
\begin{align}
  \Mset(F)=&
  \{(u_1,u_2,u_3)| u_1=u_2\le0\}
  \cup\{(u_1,u_2,u_3)| u_1=0\le u_2\}\nn
  &\cup\{(u_1,u_2,u_3)| u_2=0\le u_1\}. 
\end{align}
This is an intersecting point of 
three half planes. 
The points of $(A_1,A_1,A_1)$ form a submanifold of codimension 2
which is a curve when $\dim U=3$. 

When $k=4$, $F$ is right equivalent 
at the minimum points 
to
\begin{align}
  &(A_1,A_1,A_1,A_1)\nn
  &=
  (x_1^2+y_1^2+u_1,x_2^2+y_2^2+u_2,x_3^2+y_3^2+u_3,x_4^2+y_4^2)
  \nn
  &\qquad
  \text{at $((0_1,0_2,0_3,0_4),0)\in (\R^2)^{(4)}\times\R^3$}. 
\end{align}
The minimum function is
\begin{equation}
  {\mf}(u)=\min\{u_1,u_2,u_3,0\}.
\end{equation}
The Maxwell set is
\begin{align}
  \Mset(F)=&
  \{(u_1,u_2,u_3)| 0\ge u_1=u_2\le u_3\}\nn
  &\cup
  \{(u_1,u_2,u_3)| 0\ge u_2=u_3\le u_1\}\nn
  &\cup
  \{(u_1,u_2,u_3)| 0\ge u_3=u_1\le u_2\}\nn
  &\cup
  \{(u_1,u_2,u_3)| u_2\ge u_1=0\le u_3\}\nn
  &\cup
  \{(u_1,u_2,u_3)| u_1\ge u_2=0\le u_3\}\nn
  &\cup
  \{(u_1,u_2,u_3)| u_1\ge u_3=0\le u_2\}.
\end{align}
This is an intersecting point of six pieces of surfaces. 

\begin{figure}
\centering
\resizebox{0.75\textwidth}{!}{%
  \includegraphics{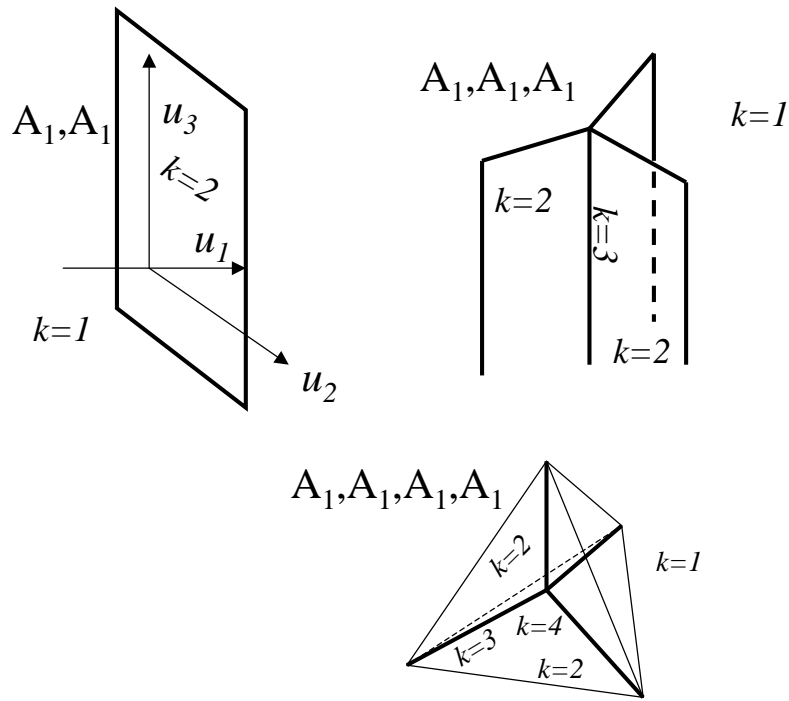}
}
\caption{Topology of the Maxwell sets for $(A_1,...,A_1)$. 
For $(A_1,A_1)$ the Maxwell set germ at $0$ 
is a surface (plane) $u_1= 0$.
For $(A_1,A_1,A_1)$, the Maxwell set is a direct product of a
Y-junction 
and $\R$. 
In the case of $(A_1,A_1,A_1,A_1)$, the Maxwell set is intersection of
six pieces of planes each of which contains an edge of the tetrahedron 
whose center is the origin. 
The numbers of minimum points at each parameters are indicated by $k$.}
\label{fig:A1k}       
\end{figure}

The embedded images for $k=2,3,4$ 
are illustrated in Figure~\ref{fig:A1k}.

\subsection{$A_3$}
\begin{figure}
\centering
\resizebox{0.75\textwidth}{!}{%
  \includegraphics{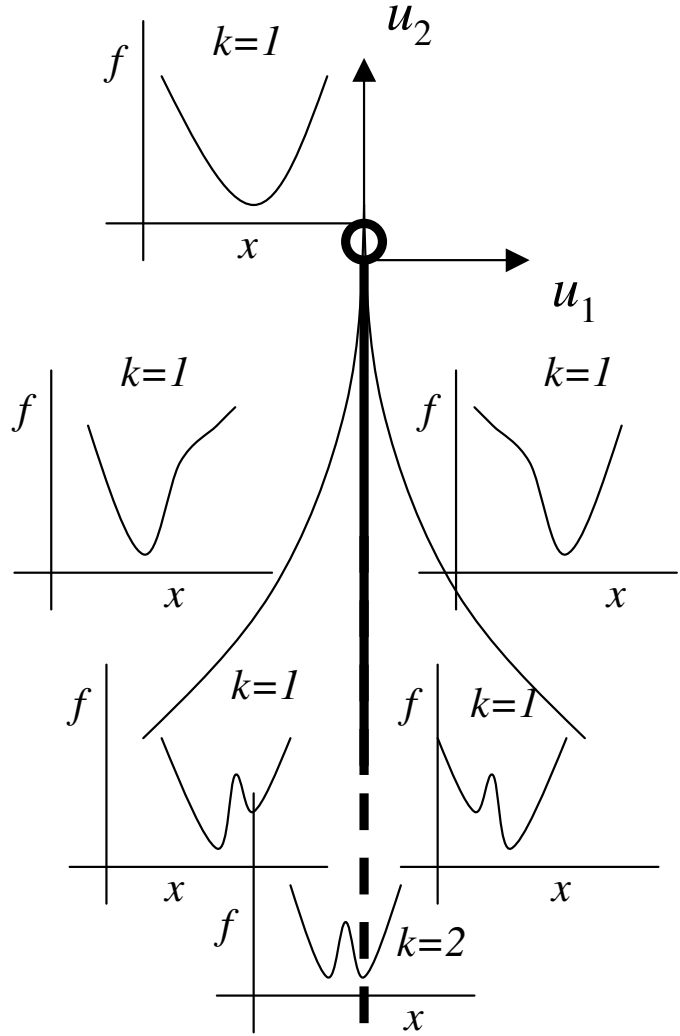}
}
\caption{Structure of $A_3$ and the Maxwell set. 
  The shape of the deformation of 
  $F (x,0,u_1,u_2)$ 
  in the parameter space $(u_1,u_2)$ is shown. 
  At the point of the Maxwell set  (bold line)
  two local minimum {\em values\/} are degenerate. On the cusp, one of 
  the two local minimum points vanishes. 
At the top of the cusp (the origin), 
two local minimum points become degenerate.
In each case, $k$ is the number of minimum points. 
The origin $A_3$ is not contained in the Maxwell set.}
\label{fig:A3}       
\end{figure}

In the case where $F$ is right equivalent 
at the minimum points 
to 
\begin{align}
  A_3(x,y,u_1,u_2)=&x^4+u_2x^2+u_1x+y^2
  \text{ at }(0,0)\in\R^{2}\times\R^3. 
\end{align}
However, there is only one {\minimal}  point
exist for $u=0$, there $u$'s in the vicinity of the origin 
where there are two {\minimal}  points $(A_1,A_1)$. 
The shape of the graph of the function $f_u(x)=F(x,0,u_1,u_2)$ changes 
as $(u_1,u_2)$ changes around the origin of $A_3$. 
This is depicted in Figure \ref{fig:A3}. 
We have 
\begin{align}
  \mf(u)=\min_x(x^4+u_2x^2+u_1x). 
\end{align}
The maxwell set germ is given by the condition of the quartic function
above having two minimum points: 
\begin{align}
  (\Mset(F),0)=(\{(u_1,u_2,u_3)\in\R^3|u_1=0,u_2<0\} ,0). 
\end{align}
The structure $A_3$ appears on the boundary of 
the surface formed by points of $(A_1,A_1)$ 
of the Maxwell set, 
where two {\minimal}  points 
$(A_1,A_1)$ become degenerate. 
The structure also appears near $(A_3,A_1)$ below. 
The point $A_3$ itself (the origin) is not contained in 
$\Mset(F)$. The structure appears at the boundary of 
$\Mset(F)$. 

\subsection{$(A_3,A_1)$}
In the case where $F$ is right equivalent 
at the minimum points 
to
\begin{equation}
(A_3,A_1)=(x_1^4+u_2x_1^2+u_1x_1+y_1^2+u_3,x_2^2+y_2^2)
\end{equation}
The minimum function is given by
\begin{align}
  \mf(u)
  =\min\{\min_{x}h(x),0\}.  
\end{align}
where $h(x):=x^4+u_2x^2+u_1x+u_3$. 
Let $x_m$ the {\minimal} point of $h(x)$. 
Then we have
\begin{eqnarray}
h'(x_m)=4x_m^3+2u_2x_m+u_1=0
\label{eq-stat-point}
\end{eqnarray}
so that
\begin{align}
  u_1&=-2x_m(2x_m^2+u_2)=:b(x_m),
  \label{eq-b}\\
  h(x_m)
  &=-3x_m^4-u_2x_m^2+u_3.
  \label{eq-hxm}
\end{align}
A point of the Maxwell set 
must satisfy either of the following:\\
(a) $h$ has two minimum points and $h(x_m)\le0$, \\
(b) $h$ has one minimum point and $h(x_m)=0$.

The case (a) is the same as the case of single $A_3$. 
The function $h$ has two minimum points if and only if 
$u_1=0$, $u_2<0$. 
The minimum value 
$h(\pm\sqrt{-u_2/2})=u_3-u_2^2/4$
must not be positive. 
Thus the part of the Maxwell set 
for the case (a)
is 
\begin{align}
  M_1=\{(u_1,u_2,u_3)|u_1=0, u_2<0, u_3\le\frac{u_2^2}4\}. 
\end{align}

Let us consider the case (b). 
If $u_1=0$, then $u_2\ge0$ must hold. The
minimum value is $h(0)=u_3$. Thus we have $u_3=0$. 
If $u_1\ne0$,
then there is always a unique solution of \Ref{eq-stat-point} 
which satisfies $u_1x_m<0$. 
This solution $x_m$ gives the unique minimum point of $h$. 
From \Ref{eq-b}, the condition $u_1x_m<0$ can be satisfied 
when 
\begin{align}
  2x_m^2+u_2>0.
  \label{eq-tmp2}
\end{align}
From $h(x_m)=0$ and
\Ref{eq-hxm}, we have
\begin{equation}
  x_m^2=\frac{-u_2\pm\sqrt{u_2^2+12u_3}}{6}.
  \label{eq-xw2}
\end{equation}
From \Ref{eq-tmp2} and \Ref{eq-xw2} (and $u_1x<0$), we have
\begin{align}
  -2u_2<\pm\sqrt{u_2^2+12u_3}> u_2. 
\end{align}
Thus the plus sign always has to be taken. 
When $u_2\ge0$, we have $u_3>0$. 
When $u_2<0$, we have $u_3>u_2^2/4$. 
In both cases, from \Ref{eq-b}, $u_1$ is expressed by $u_2$ and $u_3$
as 
\begin{eqnarray}
  u_1=\pm b\paren{\sqrt{\frac{\sqrt{u_2^2+12u_3}-u_2}{6}}}.
\end{eqnarray}
Therefore the part of the Maxwell set for the case (b) is 
\begin{align}
  M_2\setminus\{(u_1,u_2,u_3)|u_1=0,u_2<0,u_3=\frac{u_2^2}4\},
\end{align}
where
\begin{align}
  M_2=&\Bigl\{(u_1,u_2,u_3)|\nn
  &|u_1|=
  \frac{\sqrt{2}}{3\sqrt3}
  \bigparen{\sqrt{u_2^2+12u_3}+2u_2}
  \sqrt{\sqrt{u_2^2+12u_3}-u_2}
  \Bigr\}. 
\end{align}
The point $(A_3, A_1)$ is not on the boundary of the Maxwell set
according to our definition above. 

The Maxwell set is given by
\begin{align}
  \Mset(F)=M_1\cup M_2. 
\end{align}
The subset $M_2$ is diffeomorphic to a part of the bifurcation set of the
swallow-tail catastrophe. 

\begin{figure}
\centering
\resizebox{0.75\textwidth}{!}{%
  \includegraphics{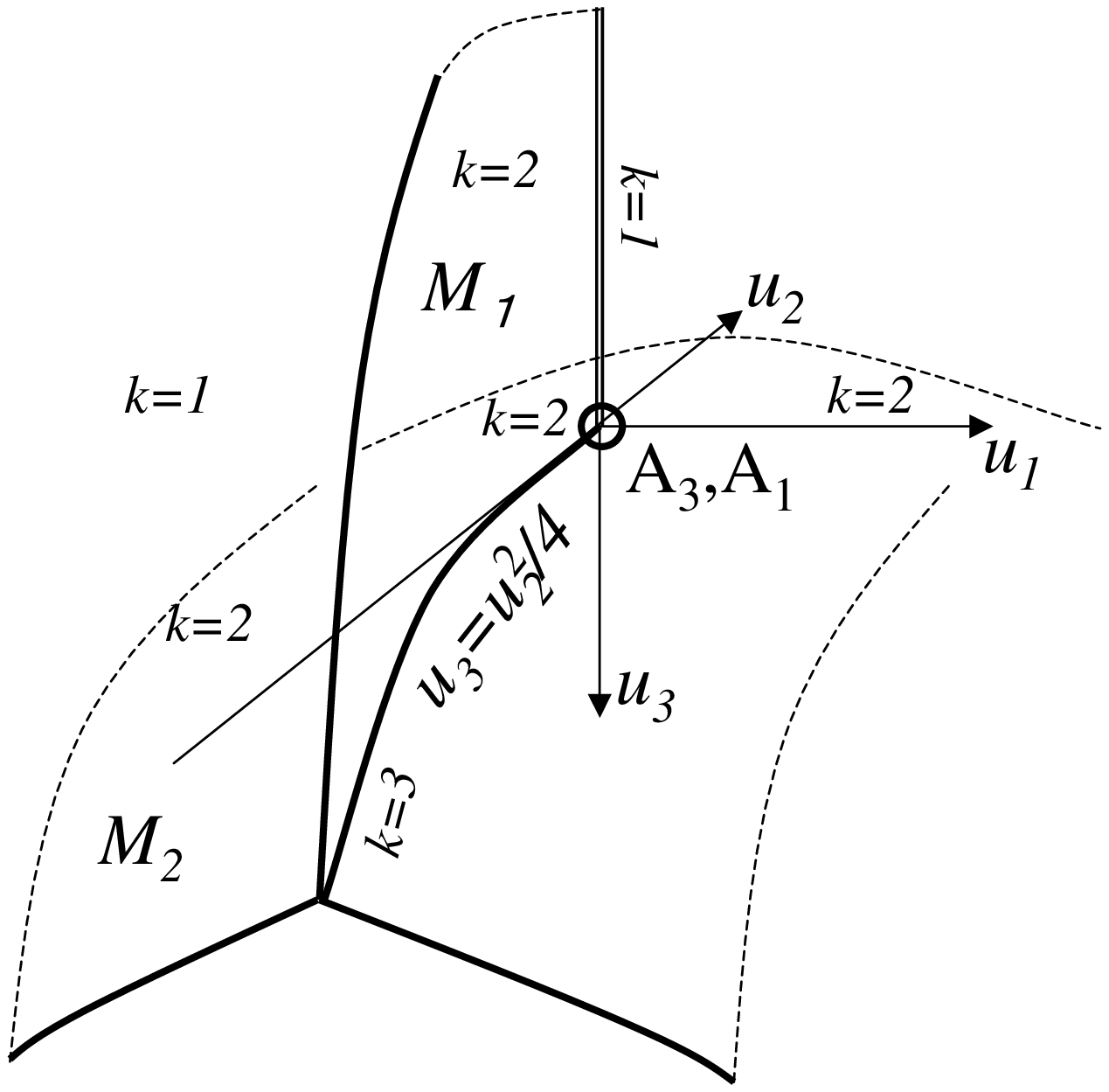}
}
\caption{Structure of $(A_3,A_1)$. 
In this case, $A_3$ competes with $A_1$.
They balance on the surface $M_2$, which is not smooth on 
$u_1=0, u_2<0$. Above the surface the
Maxwell set becomes that of $A_3$.
$k$ indicates the number of minimum points.}
\label{fig:A3A1}       
\end{figure}

The Maxwell set around $(A_3,A_1)$ is depicted in Figure \ref{fig:A3A1}.
The stable Maxwell set is diffeomorphic to the union of the broken
swallow-tail $\{x^4+u_2x^2+u_1x$ has exactly one real root$\}$ and the
 quadrant $\{u_1=0,u_3\ge u_2^2/4\}$ bounded by its line of
 self-intersection and its transversal.  
Since the boundary (the bold white line in Figure \ref{fig:A3A1}) is
the same as that of $A_3$, it has only one {\minimal}  point and is
not contained in the Maxwell set. On surfaces around $(A_3,A_1)$ 
the structure is
diffeomorphic to $(A_1,A_1)$ and the two minimum values are
degenerate. 
On the vertex of the surface $(A_1,A_1,A_1)$, three minimum values 
are degenerate.  
At $(A_3,A_1)$ these three {\minimal}   points are
degenerate into two {\minimal}  points.

\subsection{The whole structure}

We have enumerated the stable/generic local structure 
of the Maxwell set. 
Now it is easy to see that the whole of 
any stable/generic Maxwell set is 
obtained by connecting the parts in accord with 
the following rules. This shows which Maxwell sets with less minimum points
surround the Maxwell set.
\begin{equation}
  \begin{array}{cccccccc}
     & &(A_3,A_1)\\
     &\swarrow&\downarrow &\searrow\\
     (A_1,A_1,A_1)&\to&(A_1,A_1)&\leftarrow &\text{\fbox{$A_3$}}\\
     &\nwarrow&\uparrow &\\
     & &(A_1,A_1,A_1,A_1)
  \end{array}
  \label{eq-diagram-4d}
\end{equation}
An arrow means that the structure at the origin of the arrow
has the structures pointed by the arrow in the vicinity. 
The box means that the structure appears at the boundary of the
Maxwell set and the structure is not contained in the Maxwell set. 
Though the structure is not a point of the Maxwell set but corresponds
to a point of the endpoint set of the event horizon. 

In particular, we have the  following:
\begin{proposition}
The stable/generic Maxwell set does not contain its boundary. 
Any point on the boundary has the structure $A_3$. 
Accordingly, no boundary point of the stable/generic 
endpoint set $\EPS$ 
is contained in the crease set $\crset$. Any boundary point
intersects a unique null generator of $\HH$ so that 
$\HH$ is diffenentiable there. 
On the stable/generic crease set, the multiplicity of null tangent is
no more than four. 
\end{proposition}
Here a boundary point $q$ of the crease set $\crset$ (or the endpoint
set $\EPS$) is defined such that 
there is no $C^0$-submanifold of codimension 2 of $\MM$ 
which contains $q$ and which is contained entirely in $\crset$ 
(or $\EPS$). An interior point is a point at which such a submanifold
exists. 

We conclude that for stable/generic horizons,  
\begin{align}
  \crset&=\text{interior}(\EPS)=\psi(\Mset(F)),\\
  \diffset&=\text{boundary}(\EPS)
  =\psi(\overline{\Mset(F)}\cap \bifset(F)), \\
  \EPS&=\psi(\overline{\Mset(F)}),
\end{align}
where $\psi$ is the map defined in Sect.~\ref{fermat}. 
The set $\bifset(F)$ is the bifurcation set of $F$
which commonly appears in singularity theory. 
Local minimum points bifurcate on $\bifset(F)$. 

\begin{figure}
\centering
\resizebox{0.75\textwidth}{!}{%
  \includegraphics{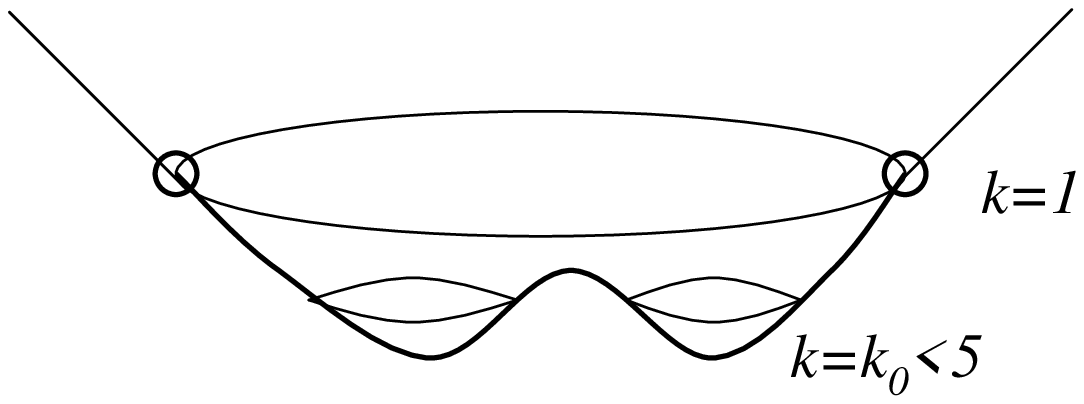}
}
\caption{A generic Maxwell set embedded into a {\spt} as a
  crease set (the spatial dimension is suppressed to two.). 
  Since its boundary is open, the crease set becomes null there.
  $k$ indicates the number of minimum points.}
\label{fig:crset}       
\end{figure}

The generic embedding of the crease set is depicted in
Fig.~\ref{fig:crset}. The endpoint set $\EPS$ of the horizon is smooth
and possesses a null tangent plane at the boundary. 
Since all structure is two-dimensional, every generic horizon admits 
time slices by tori or by higher genus surfaces. 

\begin{figure}
\centering
\resizebox{0.75\textwidth}{!}{%
  \includegraphics{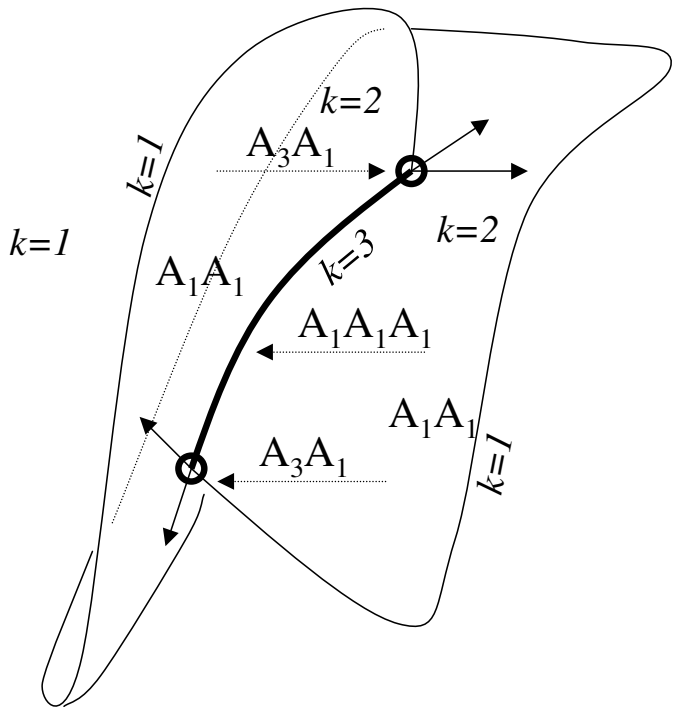}
}
\caption{An example of generic crease set. 
This is composed of two $(A_3,A_1)$, which are connected to each
other through $(A_1,A_1)$
and $(A_1,A_1,A_1)$.  
Since the boundary is that of $A_3$ and $(A_3,A_1)$, it is open. 
$k$ indicates the number of minimum points.}
\label{fig:rei}       
\end{figure}

We demonstrate an example of generic horizon.
The generic crease set is composed of the Maxwell sets above and
an example is depicted in figure \ref{fig:rei}.
They are combinations of two-dimensional segment and the
boundary is always $A_3$ which is not contained in the crease set. 
The topology of the spatial section of the horizon can be
a torus or higher genus surface or can have many components 
and its crease set has a boundary with regular null tangent plane.

\section{Other {\spt} dimensions}
\label{other-dim}
So far we have assumed the {\spt} is four-dimensional. 
Now we discuss other dimensions. 
By the same line of argument as we have presented, 
for any dimension, the equivalence of the stability and genericity
holds. Up to six {\spt} dimensions, 
all Maxwell set germs are enumerated by 
combination of finite elementary
catastrophes.  
For higher dimensions, this does not hold and there will be 
parametrized family of stable Maxwell set germs. 

In a three-dimensional {\spt}, $U$ is two-dimensional. 
The stable Fermat potential unfoldings  
and the Maxwell set germs 
are those found in the previous sections which include
at most two parameters $u_1$ and $u_2$. 
Then there are only two structures. 
One is $(A_1,A_1)$. 
The Maxwell set is given by
\begin{align}
  \Mset(F)=\{(u_1,u_2)|u_1=0\}.
\end{align}
The other is $(A_1,A_1,A_1)$. 
The Maxwell set is given by
\begin{align}
  \Mset(F)=&
  \{(u_1,u_2)| u_1=u_2\le0\}
  \cup\{(u_1,u_2)| u_1=0\le u_2\}\nn
  &\cup\{(u_1,u_2)| u_2=0\le u_1\}. 
\end{align}
The connection rule is given by
\begin{equation}
  \begin{array}{cccccccc}
  (A_1,A_1,A_1)&\rightarrow&(A_1,A_1)&
  \leftarrow &\text{\fbox{$A_3$}}
  \end{array}
\end{equation}
Thus only possible structure of the endpoint set $\EPS$ 
is a binary tree. 
In particular, the boundary of $\EPS$ does not have an intersection
with the crease set $\crset$ and
consists of the points of multiplicity one.  
The multiplicity of the endpoints of the horizon 
does not exceed three.


In five dimensions, where $U$ is four-dimensional, 
the stable Maxwell set germs include 
the direct product of $\R$ and each 
stable Maxwell set germ in a four-dimensional {\spt}.
New types of the stable Fermat potential unfolding 
$F$ emerging in five-dimensional {\spt}  are the following three cases
(Remark~\ref{rem:arn}). 

(1) 
The unfolding $F$ is right equivalent at the minimum points 
to 
\begin{align}
  &(A_1,A_1,A_1,A_1,A_1)\nn
  &=(x_1^2+y_1^2+u_1,...,x_4^2+y_4^2+u_4,
  x_5^2+y_5^2) \nn
  &\quad \text{ at } 
  ((0_1,...,0_5),0)
  \in(\R^{2})^{(5)}\times\R^3. 
\end{align}
The minimum function is 
\begin{equation}
  {\mf}(u)=\min\{u_1,u_2,u_3,u_4,0\}.
\end{equation}
The Maxwell set is 
\begin{align}
\Mset(F)=&\bigparen{\bigcup_{1\le i<j\le 4}M_{ij}}
\cup\bigparen{\bigcup_{1\le i\le 4}N_{i}},
\end{align}
where
\begin{align}
  M_{ij}&=\{(u_1,...,u_4)| u_i=u_j\le \text{the other $u_l$'s}, 0\},\\
  N_{i}&=\{(u_1,...,u_4)| u_i=0\le \text{the other $u_l$'s }\}.
\end{align}

(2) 
The unfolding $F$ is right equivalent at the minimum points 
to 
\begin{align}
  &(A_3,A_1,A_1)\nn
  &=(x_1^4+u_2x_1^2+u_1x_1+y_1^2+u_3,x_2^2+y_2^2+u_4,x_3^2+y_3^2)\nn
  &\quad
  \text{ at }
  ((0_1,0_2,0_3),0)
  \in(\R^{2})^{(3)}\times\R^3. 
\end{align}
The minimum function is given by
\begin{align}
  \mf(u)
  =\min\{\min_{x}h(x),u_4,0\}.  
\end{align}
where $h(x):=x^4+u_2x^2+u_1x+u_3$. 
In the region $u_4\ge0$ the Maxwell set is the same as that of 
$(A_3,A_1)$. 
In the region $u_4\le0$ it is the same as that of $(A_3,A_1)$
but $u_3$ is replaced by $u_3-u_4$. 
We have
\begin{align}
  &\Mset(F)\nn
  &=
  \{(u_1,u_2,u_3,u_4)|u_1=0, u_2<0, u_3\le\frac{u_2^2}4,u_4\ge0\}\nn
  &\cup
  \Bigl\{(u_1,u_2,u_3,u_4)\Bigr\vert
  |u_1|=
  \frac{\sqrt{2}}{3\sqrt3}
  \bigparen{\sqrt{u_2^2+12u_3}+2u_2}\nn
  &\qquad\times
  \sqrt{\sqrt{u_2^2+12u_3}-u_2},u_4\ge0
  \Bigr\}\nn
  &\cup
  \{(u_1,u_2,u_3,u_4)|u_1=0, u_2<0, u_3-u_4\le\frac{u_2^2}4,u_4\le0\}\nn
  &\cup
  \Bigl\{(u_1,u_2,u_3,u_4)\Bigr|
  |u_1|=
  \frac{\sqrt{2}}{3\sqrt3}
  \bigparen{\sqrt{u_2^2+12(u_3-u_4)}+2u_2}\nn
  &\qquad\times
  \sqrt{\sqrt{u_2^2+12(u_3-u_4)}-u_2},u_4\le0
  \Bigr\}
\end{align}

(3) 
The Unfolding $F$ is right equivalent at the minimum points 
to 
\begin{align}
  &A_5(x,y,u_1,...,u_4)\nn
  &=x^6+u_4x^4+u_3x^3+u_2x^2+u_1x+y^2 \nn
  &\quad \text{ at } 
  ((0_1,...,0_5),0)
  \in(\R^{2})^{(5)}\times\R^3. 
\end{align}
The minimum function is
\begin{align}
  \mf(u)&=\min h(x),\\
  h(x)&=
  x^6+u_4x^4+u_3x^3+u_2x^2+u_1x. 
\end{align}
We obtain the Maxwell set by considering 
the function $h(x)$ to have two or more global minimum points.
The condition is that $h(x)$ must be of the form
\begin{align}
  h(x)&=(x-\alpha)^2(x-\beta)^2(x^2+ax+b)+c,\\
  &a^2-4b\le0,\quad
  a,b,c\in\R. 
\end{align}
Because $h(0)=0$ and $h(x)$ does not have a $x^5$-term , 
we have 
\begin{align}
  c=-b\alpha^2\beta^2,\quad a=-2(\alpha+\beta). 
\end{align}
Thus, setting $s=\alpha+\beta$ and $p=\alpha\beta$, 
we have
\begin{align}
  h(x)&=(x^2-sx+p)^2(x^2+2sx+b)-b\alpha^2\beta^2.
\end{align}
From this equation we obtain a parametric expression for the 
Maxwell set:
\begin{align}
  \Mset(F)=\{(u_1,...,u_4)|
  &{u_1}=-2bps + 2p^2s,\nn
  &{u_2}=2bp + p^2 + bs^2 - 4ps^2,\nn
  &{u_3}=-2bs + 2ps + 2s^3,\nn
  &{u_4}=b + 2p - 3s^2,\nn
  &4p\le s^2\le b\}.
\end{align}

The connection rule is given by 
the union of the diagram \Ref{eq-diagram-4d}
and the following ones. 
Below only connections from the new three types of Maxwell sets
are shown. 
\begin{equation}
  \begin{array}{cccccccc}
     (A_3,A_1)&\leftarrow &(A_3,A_1,A_1)&\to& \text{\fbox{$A_3$}}\\
     & \swarrow&\downarrow &\searrow\\
     (A_1,A_1,A_1,A_1)&&(A_1,A_1,A_1)&&(A_1,A_1)\\
     &\nwarrow&\uparrow &\nearrow\\
     & &(A_1,A_1,A_1,A_1,A_1)
  \end{array}
  \label{eq-diagram-5d}
\end{equation}
\begin{equation}
  \begin{array}{cccccccc}
    &&(A_1,A_1)\\
    & &\uparrow\\
     (A_1,A_1,A_1)&\leftarrow & A_5&\to& (A_3,A_1)&\\
     & &\downarrow &&\\
     & &\text{\fbox{$A_3$}}\\
  \end{array}
\end{equation}
Again, the boundary is always $A_3$. 
The boundary of a generic $\EPS$ consists of points of multiplicity
one. The multiplicity on $\EPS$ does not exceed five.

\section{Conclusion and discussions}
\label{conc}
In this paper, we relate the crease set of the event horizon 
to the Maxwell set of a function unfolding 
through extended Fermat's principle. 
We have shown equivalence of stability and genericity of the
Maxwell set.
We have classified the stable Maxwell set, hence the crease set of the
horizon, 
for {\spt}s of dimension 3, 4, and 5. 
We have enumerated the parts to construct a stable crease set and 
have shown how we can connect the parts. 
In particular, the multiplicity is always one on the boundary the
endpoint set and the multiplicity of the endpoint is less than 
or equal to the dimension of the {\spt}. 

In a four-dimensional {\spt}, 
all generic horizon is possible to be realized as a
toroidal or higher genus one, 
since allowed structure is two-dimensional. Furthermore, the number of
null generators which belong to a point 
of the crease set is determined. It is concluded that the generic
crease set does not contain any of its boundary points. 


In some cases or purposes, it may be 
useful or necessary to treat the stability and the genericity under
some exact symmetry.  
Such a symmetry restricts the space of deformations of $F$ 
and our classification here gives only a sufficient conditions 
for stability or genericity in the new function space. 
An example is 
numerically generated event horizons in the study of gravitational
collapse where some exact symmetry such as axial one is imposed. 
Another example is an black hole in the brane universe scenario, where
the five-dimensional {\spt} has a exact ${\Z}_2$ symmetry. 
On those situations, the framework of this paper would require some
modifications. The classification of Maxwell sets under exact 
symmetries is our future problem. 
\section*{Acknowledgments}
We are greatly indebt to 
Professor Takuo Fukuda for 
detailed advices on mathematical concepts.

%

\begin{thebibliography}{99}
\bibitem{OW}
S. W. Hawking, {\sl Commun. Math. Phys. \bf 25 \rm (1972) 152},
P. T. Chrusciel and R. M. Wald, {\sl Class. Quant. Grav. \bf 11\rm
  (1994) L147}, 
D. Gannon,{\sl Gen. Relativ. Gravit. \bf 7 \rm (1976) 219},
J. L. Friedmann, K. Schleich and D. M. Witt, {\sl Phys. Rev. 
Lett. \bf 71 \rm (1993) 1486},
T. Jacobson and S. Venkataramani, {\sl Class. Quantum Grav.
 \bf 12 \rm (1995) 1055},
 S. Browdy and G. J. Galloway, {\sl J. Math. Phys. \bf 36
 \rm (1995) 4952},
 G.J. Galloway, K. Schleich, D.M. Witt, E. Woolgar, {\sl Phys.Rev. \bf
   D60 \rm 104039,  (1999) }  
 \bibitem{ONW}
S. A. Hughes, C. R. Keeton, P. Walker, K. Walsh, S. L. 
Shapiro and S. A. Teukolsky, {\sl Phys. Rev.  \bf D49 
 \rm (1994) 4004}, A. M. Abrahams, G. B. Cook, S. L. Shapiro and
S. A. Teukolsky 
 {\sl Phys. Rev. \bf D49 \rm (1994) 5153}, S. L. Shapiro, S. A. Teukolsky and
 J. Winicour {\sl Phys. Rev. \bf D52 \rm  (1995) 6982},
 P. Anninos, D. Bernstein, S, Brandt, J. Libson, J. 
Mass\'{o}, E. Seidel, L. Smarr, W. Suen, and P. Walker, {\sl Phys. Rev. Lett.
 \bf 74
 \rm (1995) 630},
 S. Husa and J. Winicour,{\sl Phys.Rev.\bf D60 \rm 084019  (1999)} 
\bibitem{MS1}M. Siino {\sl Phys. Rev. \bf D58 \rm 104016  (1998)}
\bibitem{MS2}M. Siino {\sl Phys. Rev. \bf D59 \rm 064006  (1999)}
\bibitem{HE}S. W. Hawking and G. F. R. Ellis, {\sl The large scale 
structure of space-time } Cambridge University Press, New York, 1973
\bibitem{BK}J. K. Beem and A. Kr\'{o}lak, {\sl J. Math. Phys.} 
  {\bf 39} (1998)
6001--6010
\bibitem{Chr98CQG}
P.~T. Chru\'siel,
\newblock {\em Class. Quantum Grav.} {\bf 15}  (1998) 3845.
\bibitem{KS}T. Koike and M. Siino, forthcoming article
\bibitem{PS}T. Poston and J. N. Stewart, {\sl Catastrophe Theory and
    its Applications, \rm Pitman, 1978} 
\bibitem{Kossowski}M. Kossowski, {\sl J. London Math. Soc.  (2) \bf 40
    \rm 179--192  (1989)} 
\bibitem{BS}M. Golubitsky and V. Guillemin, {\sl Stable mappings and
    their singularities, Graduate Texts in Math., \bf 14, \rm
    Springer-Verlag, New York-Heidelberg} 
\bibitem{MA}J. N. Mather, {\sl Advances in Math. \bf 4 \rm 301--336  (1970)}
\bibitem{ND}J. N. Mather, {\sl The nice dimensions, Lecture notes in
    Mathematics {\bf 192}, Springer-Verlag, Berlin, 207--253  (1971)} 
\bibitem{STF}R. Benedetti and J.-J. Rishler, Real algebraic sets, {\sl
    Actualit\'{e}s Mathematiques, \rm Hermann, \'{E}diteurs des
    Sciences et des Arts  (1990)}, 
C. G. Gibson, K. Wirthm\"{u}ller, A. A. du Plessis and
E. J. N. Looijenga, Topological stability of smooth mappings, {\sl
  Lecture Notes in Mathematics \bf 552, \rm Springer-Verlag, Berlin
   (1976)}, 
Y. C. Lu, Singularity Theory and an Introduction to Catastrophe
Theory, {\sl Universitext, \rm Springer-Verlag, Berlin  (1976)}, 
H. Whitney, {\sl Annals of Mathematics, \bf 81, \rm 496--549  (1965)} 
\bibitem{AN}V.I. Arnold in {\sl Dynamical  systems VIII, Encyclopedia
    of Mathematical Science Vol. 39 \rm Springer-Verlag, Chap 2,
    Sect. 3} 
\end{thebibliography}
%

\end{document}